\documentclass[12pt]{iopart}
\usepackage{iopams}

\usepackage[latin1]{inputenc}
\usepackage{graphicx}
\usepackage{amssymb}
\usepackage{color}
\usepackage{float}
\expandafter\let\csname equation*\endcsname\relax 
\expandafter\let\csname endequation*\endcsname\relax
\usepackage{amsmath}
\usepackage{amsfonts}
\usepackage{dcolumn}
\usepackage{hyperref}

\usepackage{color}

\begin{document}

\title{Stability and existence analysis of static black holes in pure Lovelock theories}

\author{Radouane Gannouji$^{1}$, Naresh Dadhich$^{2,3,1}$}
\address{$^{1}$ Astrophysics and Cosmology Research Unit, School of Mathematics, Statistics and Computer Sciences, University of KwaZulu-Natal, Private Bag X54001, Durban 4000, South Africa}
\address{$^{2}$ Centre for Theoretical Physics, Jamia Millia Islamia, New Delhi-110025, India}
\address{$^{3}$ Inter-University Centre for Astronomy \& Astrophysics, Post Bag 4, Pune 411 007, India}

\begin{abstract}
In this paper, we study existence and stability of static black holes in Lovelock theories with a particular focus on pure Lovelock black holes. We derive the equation of stability from action without using S-deformation approach. It turns out that though pure-Lovelock black hole in even dimension is always unstable, however introduction of $\Lambda$ stabilizes it by prescribing a lower threshold mass while there also exists an upper bound on mass which is given by existence of horizon. We also study stability of dimensionally continued black holes as well as of pure Lovelock analogue of BTZ black hole in all odd dimensions. 
\end{abstract}


\section{Introduction}

It is well known that string theory which is proclaimed as a theory of everything can only live in higher dimensions and one of the key issues, besides unification of all forces, it is supposed to address is quantum gravity. Hence higher dimensional probing of gravity is quite in line with the current work in fundamental physics. The most natural higher dimensional generalization of Einstein gravity is Lovelock polynomial gravity \cite{lov} which includes linear order Einstein gravity for $N=1$ and quadratic Gauss-Bonnet (GB) for $N=2$, where $N$ is the degree of homogeneous polynomial in Riemann curvature. This is the unique generalization which retains the second order character of the equation of motion, and $N$th order term makes non-zero contribution to the equation only in dimension $d\geq 2N+1$. It is therefore truly a higher dimensional generalization. It may be noted that among others, quadratic GB term has been shown to be relevant to low energy limit of string  theory \cite{Gross:1986iv,Zumino:1985dp,Zwiebach:1985uq}. This may perhaps be indicative of the fact that high energy limit of classical gravity may be Lovelock gravity. This is how one of us has argued \cite{d1, d2} for quite some time on purely classical motivation for higher dimensions. For inclusion of high energy effects within classical framework we should involve higher orders of Riemann curvature and yet if we demand the equation to remain second order, we are uniquely led to Lovelock polynomial and higher dimensions \cite{d2}. This is how Lovelock gravity may perhaps arise as high energy limit of classical gravity and hence it may serve as an  intermediatory state between classical and quantum gravity \cite{d3}. Thus notwithstanding the strong string theory motivation for higher dimensional study of gravity, the point we would like to make is that there is also equally pertinent classical motivation.

Further it has been argued that pure Lovelock gravity imbibes one of the universal features of gravity that it is kinematical in all odd dimensions like Einstein gravity is in $3$ dimensions \cite{dgj1}. What it means is that $R^{(N)}_{ab}=0$ implies $R^{(N)}_{abcd}=0$ in all $d=2N+1$ dimensions. Here $R^{(N)}_{abcd}$ is defined such that vanishing of trace of its Bianchi derivative gives the same divergence free second rank symmetric tensor as that obtained from variation of $N$th order Lovelock polynomial action \cite{bianchi}. It also turns out that pure Lovelock $\Lambda$-vacuum solution asymptotically goes over to corresponding Einstein Schwarzschild-dS/AdS solution even though action was free of Einstein term \cite{dpp1}. Further there is thermodynamical universality for pure Lovelock static black holes, for instance entropy in even $d=2N+2$ dimension would always go as $A^{1/N}$ where $A$ is area of horizon \cite{dpp2}. Therefore a very strong case has been made \cite{prague} for pure Lovelock equation,
\begin{align}
G^{(N)}_{ab} = -\kappa T_{ab} - \Lambda g_{ab},
\end{align}
which includes only one $N$th order term. There exists general solution for vacuum of this equation describing static black hole \cite{dpp1}. Among other general features of Lovelock gravity, very recently it has also been shown that bound orbits around a static black hole exist in all even $d=2N+2$ dimensions \cite{dgj2}.

It should be noted that pure Lovelock equation is a classical gravitational equation relevant for $d=2N+1, 2N+2$ dimensions for a given $N$. It is not a correction arising out of some other theory like string theory. This is a basic difference of viewpoint in pure Lovelock and Einstein-Lovelock theories. From this standpoint Einstein gravity is relevant only for  $d= 3, 4$ dimensions and similarly for $d=5, 6$, is Gauss-Bonnet, and hence there is no question of inclusion of Einstein or other terms $< N$. We thus have only one coupling constant for one force which would be determined experimentally as $G$ is determined for Einstein gravity. When there are more than one coupling constants, there higher order couplings arise as measure of corrections to the first order Einstein gravity. Pure Lovelock is therefore entirely on different plane.

Black holes are by far the most interesting solutions of gravitational theories. Hence their existence, uniqueness and stability are of utmost importance and have been intensely discussed in the literature. In higher dimensions vacuum solution with the usual spherical topology $S^{d-2}$ is static and unique \cite{hollands}. This is very important for black hole stability analysis. In this paper, we will focus on the study of stability of pure Lovelock black holes with horizon having spherical topology. The linear stability was studied for $4$-dimensional Einstein black holes. The metric perturbations are decomposed according to their transformation properties under two-dimensional rotations. They are classified by transformation under parity, namely odd (axial) \cite{Regge:1957td} and even (polar) \cite{Zerilli:1970se}. The two modes give rise to the same Schrodinger-type differential equation for perturbations. Finally stability of $4$-dimensional  black holes has been thoroughly investigated by several authors in \cite{Vishveshwara:1970cc,Price:1971fb,Wald:1979}.

In higher dimensions, there exists an additional tensor mode \cite{Gibbons:2002pq}. Following this, a gauge-invariant formalism was developed in \cite{Kodama:2003jz,Ishibashi:2003ap} where  perturbations are decomposed into three types of gravitational variables, depending on how they transform with respect to horizon. Hence we have three types of perturbations, tensor, vector and scalar. The last two types correspond, respectively, to odd and even modes in four-dimensional case while tensor perturbations are new and emerge only in higher dimensions. Following the scheme proposed in \cite{Kodama:2003jz,Ishibashi:2003ap}, stability of higher dimensional black holes has been an active topic of research in recent years. In the case of Einstein-Gauss-Bonnet (EGB) gravity,  stability analysis under scalar and vector perturbations was carried out in \cite{Gleiser:2005ra} and it was later generalized to any Lovelock \cite{Takahashi:2010ye}. Also tensor perturbations were studied in \cite{Dotti:2004sh,Dotti:2005sq} for EGB case, then it was  generalized to third order Lovelock in \cite{Takahashi:2009dz} and to any Lovelock order in \cite{Takahashi:2009xh}. It was shown \cite{Takahashi:2010gz} that vector perturbations are stable as long as tensor perturbations are stable. Also there exists an instability of Lovelock black holes with small mass under tensor perturbations in even-dimensions and under scalar perturbations in odd-dimensions.

It may be noted that in all odd $d=2N+1$ dimensions, there exist analogues of $3$-dimensional BTZ black hole \cite{btz}, and let us call them as pure Lovelock odd $d=2N+1$ dimensional black holes. Then $N=1$ is the BTZ black hole. It is however possible to have BTZ-like black holes in even dimensions with a non trivial geometry of the horizon \cite{Canfora:2010rh,Anabalon:2011bw}. For satability analysis we shall employ a new approach in which the stability equation directly follows from the second order of the action without the use of S-deformation method. One of the important results of our analysis is the important role played by  $\Lambda$ in imparting stability to otherwise unstable pure Lovelock black hole. This however comes at a price which makes mass of black hole bounded on either side where upper bound is dictated by existence of horizon while lower by stability.

Notice that we focus on the particular case of tensor perturbations, because we work in the critical even $2N+2$ dimensions for pure Lovelock gravity. In fact as alluded in the beginning, Lovelock gravity in odd and even dimensions is radically different \cite{dgj1,bianchi,dpp1,dpp2,prague,dgj2}. Also scalar perturbations do not provide an additional constraint for even dimensions \cite{Takahashi:2010gz} hence we focus with tensor perturbations in critical even $2N+2$ dimensions.

The plan of the paper is as follows. In Sec. 2 we give the basic Lovelock formulation which is followed by recall of static black holes solutions in Sec. 3. In Sec. 4, we present the Ishibashi-Kodama formalism and derive the master equation for tensor perturbations and discuss  stability by using S-deformation technique, and in Sec. 5 we derive the same equation by expanding action in second order which also gives us no-ghost condition without use of S-deformation. In Sec. 6, we obtain the upper bound on black hole mass in terms of 
$\Lambda$ as a condition for existence of a horizon. Next we consider stability of pure Lovelock even and odd dimensions in Sec. 7 and of dimensionally continued black holes in Sec. 8. In Sec. 9 we compare stability of Einstein, Einstein-Gauss-Bonnet, dimensionally continued and pure Lovelock black holes and the paper is rounded off by discussion.  

\section{Lanczos-Lovelock gravity}

We consider the lagrangian in d dimensions

\begin{align}
\mathcal{L}= \sqrt{-g} \sum_{m=0}^{N}\alpha_m \mathcal{L}_m\,,
\end{align}
where we define the maximum integer $N\equiv [(d-1)/2]$, $\alpha_m$ are arbitrary constants which represent the couplings and $\mathcal{L}_m$ is given by

\begin{align}
\mathcal{L}_m=\frac{1}{2^m}\delta^{a_1 \cdots a_{2m}}_{b_1 \cdots b_{2m}}R_{a_1 a_2}^{~~~~b_1 b_2}\cdots R_{a_{2m-1} a_{2m}}^{~~~~~~~~~~b_{2m-1} b_{2m}}\,,
\end{align}
where $R_{\bullet \bullet}^{~~\bullet\bullet}$ is the Riemann tensor and $\delta^{a_1 \cdots a_{2m}}_{b_1 \cdots b_{2m}}$ is the generalized Kronecker delta of order $2m$. In the following, we use $(a,b)$ as generic indices, whereas $(i,j,k,l)=(2,3,\cdots,n+1)$ and $n=d-2$.

We consider spacetime of the following form\footnote{Note that for vacuum solution, it is $R^t_t = R^r_r$ that implies $g_{tt}g_{rr}=-1$}
\begin{align}
\label{eq:metric}
{\rm d}s^2=-f(r){\rm d}t^2+\frac{{\rm d}r^2}{f(r)}+r^2\bar \gamma_{ij}{\rm d}x^i {\rm d}x^j\,,
\end{align}
with $f(r)=\kappa-r^2\psi(r)$, and $\bar \gamma_{ij}$\footnote{A bar will be added to all the tensors associated to the n-manifold} is the metric of the $n=d-2$-dimensional constant curvature space with a curvature $\kappa=1,0$ or $-1$.

The vacuum equation then reduces to solving the master algebraic equation \cite{Boulware:1985wk,Wheeler:1985qd,Deser:2005pc}
\begin{align}
\label{eq:psi}
\sum_{m=0}^N \hat{\alpha}_m \psi^m=\frac{\mu}{r^{d-1}}\,,
\end{align}
Here the coefficients $\hat{\alpha}_m$ are given by
\begin{align}
\hat{\alpha}_m = \alpha_m \frac{(d-2)!}{(d-1-2m)!}\,.
\end{align}
$\mu$ is a constant of integration related to the mass by $\mu=16\pi G M_{\text{ADM}}/\Omega_{n}$, $\Omega_{n}=2\pi^{(n+1)/2}/\Gamma((n+1)/2)$ is the area of a unit $n$-sphere ($n=d-2$), $G$ denotes the $d$-dimensional Newton constant and $M_{\text{ADM}}$ is the Arnowitt-Deser-Misner mass.

\section{Perturbations and stability}
In higher dimensions, perturbations can be decomposed into scalar, vector and tensor modes according to how they transform under $SO(d-1)$ \cite{Ishibashi:2003ap,Kodama:2003jz}. In $d=4$, a  vector perturbation corresponds to axial (odd) mode while scalar perturbation corresponds to polar (even) mode. Finally, an additional tensor mode is present in dimensions $>4$ as there are no suitable tensor harmonics in $d=4$ \cite{Higuchi:1986wu}.

Here we study tensor perturbations around solution (\ref{eq:metric}) of the form,

\begin{align}
g_{ab}\rightarrow g_{ab}+f_{ab}\,,
\end{align}
where $f_{ab}=0$ unless $(a,b)=(i,j)$, and

\begin{align}
f_{ij}(t,r,x)=r^2\phi(t,r) \bar h_{ij}(x)\,,
\end{align}
here $\bar h_{ij}$ is a TT tensor (traceless-transverse) with respect to the metric $\bar \gamma_{ij}$, solving the eigenvalue problem

\begin{align}
\bar \gamma^{ij} \bar h_{ij}=0\,, ~~ \bar \nabla^i \bar h_{ij}=0\,, ~~ \bar \Box \bar h_{ij}=\gamma \bar h_{ij}\,,
\end{align}
where $\bar \nabla^i$ is covariant derivative with respect to $\bar \gamma_{ij}$ and $\bar \Box=\bar \nabla^i \bar\nabla_i$.

Now Riemann tensor at the first order gives

\begin{align}
R_{ab}^{~~cd}\rightarrow R_{ab}^{~~cd}+\delta R_{ab}^{~~cd}\,,
\end{align}
where

\begin{align}
\delta R_{ab}^{~~cd} = \frac{1}{2}\Bigl(R_{ab}^{~~e[c}f_{e}^{~d]}-\nabla_{[a}^{~[c}f_{b]}^{~d]}\Bigr)\,,
\end{align}
from which we have

\begin{align}
\delta R_{0i}^{~~0j} &= \Bigl(\frac{\ddot \phi}{2f}-\frac{f'}{4}\phi'\Bigr)\bar h_i^j\,,\\
\delta R_{1i}^{~~1j} &= \Bigl[-f\frac{\phi''}{2}-(\frac{f'}{4}+\frac{f}{r})\phi'\Bigr]\bar h_i^j\,,\\
\delta R_{ij}^{~~kl} &= -\frac{1}{2}\Bigl[\frac{\phi}{r^2}\bar\nabla_{[i}^{~[k}\bar h_{j]}^{~l]}+(\frac{\kappa\phi}{r^2}+\frac{f\phi'}{r})\delta_{[i}^{[k}\bar h_{j]}^{~l]}\Bigr]\,.
\end{align}
From the last equation, we have the useful relation

\begin{align}
\delta_l^j\delta R_{ij}^{~~kl} = -\frac{1}{2}\Bigl[(n-2)\frac{f\phi'}{r}+\frac{\gamma-2\kappa}{r^2}\phi\Bigr]\bar h_i^k\,.
\end{align}
By using the previous formulae, it can be shown that $\delta\mathcal{G}_0^{~a}=\delta\mathcal{G}_1^{~a}=0$ and

\begin{align}
\delta\mathcal{G}_i^j&=-\sum_{m=0}^N\frac{m\alpha_m}{2^{m+1}}\delta_{ib_1\cdots b_{2m}}^{j a_1\cdots a_{2m}}\delta R_{a_1 a_2}^{~~b_1 b_2} R_{a_3 a_4}^{~~b_3 b_4}\cdots R_{a_{2m-1} a_{2m}}^{~~b_{2m-1} b_{2m}}\nonumber\\
&=-\sum_{m=0}^N\frac{m\alpha_m}{2^{m+1}}\Bigl[4\delta_{0ilb_3\cdots b_{2m}}^{0jk a_3\cdots a_{2m}}\delta R_{0 k}^{~~0 l}
+4\delta_{1ilb_3\cdots b_{2m}}^{1jk a_3\cdots a_{2m}}\delta R_{1 k}^{~~1 l}\nonumber\\
&+\delta_{ilpb_3\cdots b_{2m}}^{jkq a_3\cdots a_{2m}}\delta R_{kq}^{~~lp}
\Bigr]
R_{a_3 a_4}^{~~b_3 b_4}\cdots R_{a_{2m-1} a_{2m}}^{~~b_{2m-1} b_{2m}}
\end{align}
which after some calculations along with $\gamma^{ij}\bar h_{ij}=0$, we obtain

\begin{align}
\delta\mathcal{G}_i^j=A\Bigl(\delta R_{0i}^{~~0j}+\delta R_{1i}^{~~1j}\Bigr)+B\delta R_{ik}^{~~jl}\delta_l^k
\end{align}

where

\begin{align}
\label{eq:A}
A &= \sum_{m=0}^N m \alpha_m \frac{(n-2)!}{(n+1-2m)!}\psi^{m-2}\Bigl[(n-1)\psi+(m-1)r\psi'\Bigr]\\
B &= \sum_{m=0}^N m \alpha_m \frac{(n-3)!}{(n+1-2m)!}\psi^{m-3}\Bigl[(n-1)(n-2)\psi^2\nonumber\\
&+(m-1)(m-2)r^2\psi'^2+(m-1)r\psi\Bigl(2(n-1)\psi'+r\psi''\Bigr)\Bigr].
\end{align}
As it was noticed in \cite{Takahashi:2009xh}, the equations are simpler if we define the function $h=r^{n-2} A$. Hence we write

\begin{align}
\delta\mathcal{G}_i^j=\frac{h}{r^{n-2}}\Bigl(\delta R_{0i}^{~~0j}+\delta R_{1i}^{~~1j}\Bigr)+\frac{h'}{(n-2)r^{n-3}}\delta R_{ik}^{~~jl}\delta_l^k .
\end{align}
After setting $\phi(t,r)=e^{\omega t}\chi(r)$, which is possible because the background is static (we use the separation of variables and integrate over time), $\delta \mathcal{G}^i_j=0$ gives

\begin{align}
-f^2\chi''-\Bigl(\frac{f^2h'}{h}+\frac{2f^2}{r}+f f'\Bigr)\chi'+\frac{(2\kappa-\gamma)f h'}{(n-2)rh}\chi=-\omega^2 \chi .
\end{align}
By further introducing $\chi=\Phi/(r\sqrt{h})$ for $h>0$ (or $\chi=\Phi/(r\sqrt{-h})$ for $h<0$) and switching to ``tortoise'' coordinate $r^\ast$, defined by $dr^\ast/dr=1/f$, we can rewrite the previous equation in the Shr\"{o}dinger form with "energy" eigenvalue $E=-\omega^2$,

\begin{align}
\label{eq:Master}
-\frac{d^2\Phi}{dr^{\ast 2}}+V(r(r^\ast))\Phi=-\omega^2 \Phi\equiv E\Phi
\end{align}
where potential is given by,

\begin{align}
V(r)=\frac{(2\kappa-\gamma)f h'}{(n-2)rh}+\frac{f}{r\sqrt{h}}\frac{d}{dr}\Bigl(f\frac{d}{dr}(r\sqrt{h})\Bigr).
\end{align}
and $\mathcal{H}$ is a differential operator in the Hilbert space of square integrable functions. The solution will be perturbatively stable if and only if the differential  operator $\mathcal{H}\equiv -\frac{d^2}{dr^{\ast 2}}+V$ acting on functions defined in the region $f>0$ is a positive self-adjoint operator, so that it has no negative eigenvalues $(E<0)$. In fact let us suppose that we have an unstable mode with $\omega \in \mathbb{R}^+$ which means $E<0$, hence we have

\begin{align}
\label{eq:defi}
E\int |\Phi|^2 dr^\ast &=\int \Phi \mathcal{H} \Phi dr^\ast= \int \Bigl(|\frac{d\Phi}{dr^\ast}|^2+V|\Phi|^2\Bigr)~dr^\ast
\end{align}
where integration is defined in the region $f>0$ and we performed a partial integration.

If potential is positive, this leads to a contradiction because left-hand side is negative while right-hand side is positive. Thus mode is stable if potential is positive. In contrast, if  potential is negative, we cannot conclude anything about stability. For that, the equation has to be further analysed by using "S-deformation'' technique \cite{Ishibashi:2003ap}. This deformation is useful for transforming a partly negative potential to a positive-definite one. We have

\begin{align}
\label{eq:def}
\int (|\frac{d\Phi}{dr^\ast}|^2+V|\Phi|^2) ~dr^\ast=\int (|D \Phi|^2+W|\Phi|^2) ~dr^\ast\,,
\end{align}
where
\begin{align}
D& \equiv \frac{d}{dr^\ast}+S\,,\\
W&=V+\frac{dS}{dr^\ast}-S^2\,,
\end{align}
and $S$ is "S-deformation" function to be defined. Hence as shown in \cite{Takahashi:2009xh} and
by choosing

\begin{align}
S=-\frac{d}{dr^\ast}\ln(r\sqrt{h})\,,
\end{align}
gives

\begin{align}
W=\frac{(2\kappa-\gamma)f h'}{(n-2)rh}\,.
\end{align}
The "new" potential depends linearly on the factor $2\kappa-\gamma$ which is positive and can be very large. For example when $\kappa=1$, we have $2\kappa-\gamma=l(l+n-1)$ which can be sufficiently large for high harmonics $(l)$.

In the case where $h'/h>0$ deformed potential $W$ is positive which implies stability of the solution. While if $h'/h<0$ potential can be sufficiently negative for high harmonics and therefore eqn. (\ref{eq:def}) will be negative. Hence there is existence of unstable modes with negative energy states ($\omega>0$). Because potential can be as negative as desired, we can always find a negative mode solution that is normalizable. A sufficient condition of the instability is $\int W {\rm d}r<0$ \cite{Buell}, which is trivially satisfied for high harmonics. Thus stability can be read from the sign of deformed potential $W$ which implies that the spacetime is stable iff $h'/h>0$ on the domain of definition $f>0$.

\section{No ghost condition}

In this section, we adopt another approach, by expanding action at second order of perturbations. It is more convenient from a computational point of view and gives easily the conditions for the avoidance of ghosts and Laplacian instabilities. The approach has been widely used for the stability analysis of various spacetimes, starting with the seminal work for the Schwarzschild spacetime \cite{Moncrief:1974am}, see also the same approach in the context of modified gravity \cite{DeFelice:2011ka} or cosmology \cite{Maldacena:2002vr}.  The relevant formulae are given in Appendix. After some lengthy calculations, we have for action on shell

\begin{align}
\mathcal{S}&=\int  {\rm d}^dx \sqrt{\bar \gamma}~   \frac{r^2 h}{4f}\Bigl[\dot\phi^2-f^2\phi'^2-W\phi^2\Bigr] \bar h_i^{~j} \bar h_j^{~i}\,,
\end{align}

From which, we can derive no ghost condition as $h>0$. Hence a ghost mode is always present around  spherically symmetric static background in vacuum if $h<0$. However existence of ghost does not necessarily imply instability if mode is massive enough. In fact, theory might be considered as valid below some cutoff scale and should be completed at higher energies. Therefore if mass of ghost mode is larger than a cutoff scale $M_{\text{cutoff}}$ (let us say $M_{\text{Pl}}$),  instability can be disregarded. But in this particular case, we will have instability at least for monopole perturbation $l=0$ which is massless. Hence we will also consider an additional condition $h>0$.

Notice that $W$ appears as effective mass squared. Hence we can deduce positivity of $W$ without using "S-deformation". Defining new variable $\phi=\Phi/(r\sqrt{h})$, and switching to the tortoise coordinate, we have
\begin{align}
\mathcal{S}&=\frac{1}{4}\int  {\rm d}^n x  {\rm d}t  {\rm d}r^\ast \sqrt{\bar\gamma}   \Bigl[\Bigl(\frac{d\Phi}{dt}\Bigr)^2-\Bigl(\frac{d\Phi}{dr^\ast}\Bigr)^2-V\Phi^2\Bigr] \bar h_i^{~j} \bar h_j^{~i}\,,
\end{align}
from which eqn. (\ref{eq:Master}) can be easily derived.

\section{Pure Lovelock}

In the following, we will consider pure Lovelock vacuum solution \cite{Cai:2006pq} for which we have

\begin{align}
\hat{\alpha}_{N} \psi^{N}+\hat{\alpha}_0=\frac{\mu}{r^{d-1}}\,,
\end{align}
for a fixed $N$. So we have the solution,

\begin{align}
\label{eq:BH}
f(r)=\kappa\pm r^2\Bigl(\frac{M}{r^{d-1}}+\Lambda\Bigr)^{1/N}\,,
\end{align}
where we have defined $M=\mu/\hat\alpha_N$ and $\Lambda=-\hat\alpha_0/\hat\alpha_N$. There are 2 families of solutions corresponding to the sign in the metric function. The positive branch exists when $N$ is even while the negative branch exists for all dimensions.

\subsection{Pure Lovelock in even dimensions}

In this case, $d=2N+2$, we can rewrite the solution (\ref{eq:BH}) as
\begin{align}
\label{eq:BH2}
f(r)=\kappa\pm r^2\Bigl(\frac{M}{r^{2N+1}}+\Lambda\Bigr)^{1/N}\,,
\end{align}
The positive mass solution has a central spacelike (timelike) singularity for the negative (positive) branch.

\subsubsection{$\Lambda=0$ case:}
In this case, we need to assume positivity of mass $M\ge 0$. The spacetime with $M=0$ corresponds to  Minkowski (hyperbolic) spacetime for $\kappa=1$ ($\kappa=-1$). There is no horizon when $\kappa=0,1$ for positive branch or when $\kappa=-1$ for negative branch. The solution represents  spacetime with a globally naked singularity. When $\kappa=-1$ for positive branch, there is a cosmological horizon at $r=M$. Therefore this solution represents a spacetime with a globally naked singularity, while there is an event horizon at $r=M$ for negative branch with $\kappa=1$.

\subsubsection{$\Lambda<0$ case:}
For a well-defined theory, the condition $M> 0$ should be satisfied except for $d=4$ or $N=1$. Besides central singularity at $r=0$, there also occurs a branch singularity at $r=r_b=(-M/\Lambda)^{1/(2N+1)}$ as in the Einstein Gauss-Bonnet gravity \cite{Torii:2005xu}. The metric is finite at $r=r_b$ but its derivative blows up and it  becomes complex when $r>r_b$. Around branch singularity, the Kretschmann scalar behaves as $(r_b-r)^{(1-2N)/N}$ and singularity is timelike, null or spacelike for $\kappa=1,0,-1$ respectively.

For $\kappa=0, 1$, for positive branch and $\kappa=-1$ for negative branch, the solution has no horizon and hence represents the spacetime with a globally naked singularity.

\begin{figure}
\begin{center}\includegraphics[scale=.6]{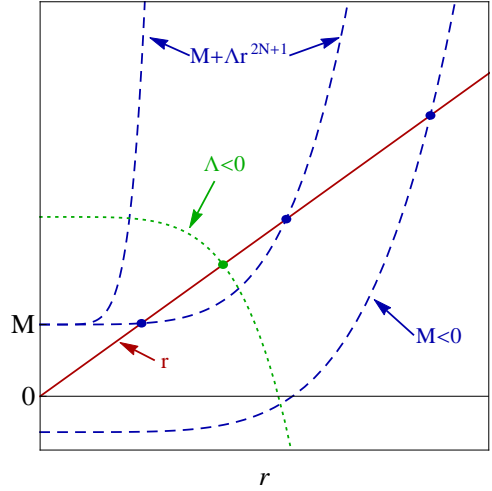}\end{center}
\caption{Solutions of the equation $r=M+\Lambda r^{2N+1}$ in the case $\Lambda>0$ (dashed line) and $\Lambda<0$ (dotted line). When $M>0$, we have 0 or 2 horizons depending on the values of the parameters while when $M<0$ we have always one horizon. We see that for $M>0$, the horizons are at $r>M$. For $\Lambda<0$, we have always one positive solution at $r<M$.}
\label{fig:hor}
\end{figure}

For negative branch and $\kappa=1$, the solution has one event horizon at $r<M$ (see Fig.\ref{fig:hor}) while for positive branch and $\kappa=-1$, there is one cosmological horizon at $r<M$. The latter case represents the spacetime with a naked singularity. Notice that branch singularity is always outside the horizon.

\subsubsection{$\Lambda>0$ case:}
If $M=0$, it is de Sitter spacetime. If mass is negative, it is defined only for $r>r_b$, where $r_b$ is location of  branch singularity. If further $\kappa=0$, the horizon coincides with the singularity and for $\kappa=1, -1$ there is a naked singularity for positive (negative) branch. The negative branch with $\kappa=1$ has one horizon (see Fig.\ref{fig:hor}), but $f(r_b)=1$ hence  branch singularity lies inside cosmological horizon ($f'(r_\text{horizon})<0$). For positive branch and $\kappa=-1$, we have an event horizon containing branch singularity.

When $M>0$ only central singularity exists. It is naked singularity for $\kappa=0$ and for  positive (negative) branch with $\kappa=1, -1$. While the situation is more complicated for  positive (negative) branch and $\kappa=-1, 1$. In fact horizons are located at $f(r)=0$ which gives
\begin{align}
\label{eq:hor}
r=M+\Lambda r^{d-1} = M+\Lambda r^{2N+1}.
\end{align}
Hence we have 0 or 2 real positive solutions of the previous equation as it can be seen from the Fig.\ref{fig:hor}.\\
In the case where there is no solution, we have a naked singularity. The second situation has 2 positive real roots of the equation (\ref{eq:hor}), the smallest root is an event horizon (inner horizon) for the negative (positive) branch and $\kappa=1$ ($\kappa=-1$) while the second root is a cosmological (event) horizon. 

The existence of the horizons depend on the parameters of the model. $(\Lambda,M)$ should be small enough in order to have a black hole. We notice that (\ref{eq:hor}) is a polynomial of order $d-1$ for which the discriminant $(\Delta)$ can be calculated. Contrary to the quadratic case, the sign of the discriminant will not be relevant. But when $\Delta=0$, at least 2 roots will be equal either real or not. In our particular case, the 2 real roots will be equal when $\Delta=0$. Hence the discriminant will give us the conditions of existence of the horizons. We can calculate directly the discriminant which is a determinant of an $(2d-3)\times(2d-3)$ Sylvester matrix or we can use the standard theorems of the resultant (res) of the polynomials $(P,P')$ where $P\equiv \Lambda r^{d-1}-r+M$, we have

\begin{align}
\Lambda (-1)^{(d-1)(d-2)/2}\Delta &= \text{res}_{d-1,d-2}(P,P')=\text{res}_{d-1,d-2}(P+QP',P')\,, \quad Q\equiv -\frac{r}{d-1}\nonumber\\
&=(-1)^{(d-2)^2}\Bigl(\Lambda (d-1)\Bigr)^{d-2} \text{res}_{1,d-2}\Bigl(\frac{2-d}{d-1}r+M,P'\Big)\nonumber\\
&=\Bigl[\Lambda (1-d)\Bigr]^{d-2} \Bigl(\frac{2-d}{d-1}\Bigr)^{d-2}P'\Bigl(\frac{d-1}{d-2}M\Bigr)\nonumber\\
&=\Bigl[\Lambda (d-2)\Bigr]^{d-2}\Bigl[\Lambda M^{d-2}\frac{(d-1)^{d-1}}{(d-2)^{d-2}}-1\Bigr]\,.
\end{align}
Which gives the condition of the existence of 2 horizons

\begin{align}
\label{existence}
\Lambda M^{d-2}<\frac{(d-2)^{d-2}}{(d-1)^{d-1}}\,.
\end{align}
Hence we found an extreme mass $M_{\text{ex}}=(d-2)/(\Lambda^{1/(d-2)} (d-1)^{(d-1)/(d-2)})$. We have 2 or 0 horizons for $M<M_{\text{ex}}$ and $M>M_{\text{ex}}$ respectively.\\
Notice that we have also a maximum mass for Einstein case in all dimensions if $\Lambda>0$ and $\kappa=1$, then 
\begin{align}
\mu<2^{\frac{5-d}{2}} \frac{d-2}{d-1}\Bigl[(d-2)(d-3)\Bigr]^{\frac{d-3}{2}}\Lambda^{\frac{3-d}{2}}\,.
\end{align}
Finally, if $M=M_{\text{ex}}$, the 2 solutions coincide and the horizon is degenerate. It represents the extreme black hole spacetime. For this solution, we have $f'(r_H)=0$ at the horizon ($r_H$) and
\begin{align}
f'(r_H)=\pm\frac{2}{r^2}\Bigl(r-M\frac{d-1}{d-2}\Bigr)\,,
\end{align}
which gives easily the position of the horizon $r_H=M (d-1)/(d-2)$. Hence we have an additional condition on the position of the horizons in the case where $M<M_{\text{ex}}$, the first horizon is at $M<r<M (d-1)/(d-2)$ and the outer horizon is at $r>M (d-1)/(d-2)$.

\subsection{Pure Lovelock in odd dimensions}

In the following, we will consider pure Lovelock vacuum solution odd dimensions $d=2N+1$ for which we have

\begin{align}
\label{eq:BH3}
f(r)=\kappa\pm r^2\Bigl(\frac{M}{r^{2N}}+\Lambda\Bigr)^{1/N}\,,
\end{align}

As previously, the positive branch exists when $N$ is even while the negative branch exists for all $N$. Notice that the spacetime is regular, we do not have a central singularity. But we can have existence of horizon and branch singularity. The discussion about branch singularity follows the same idea than in even dimensions and we will not reproduce it here. The existence of the horizon demands $k=1$ for the negative branch and $k=-1$ for the positive branch. Considering the equation $f(r)=0$, we have a unique solution $r_H=((1-M)/\Lambda)^{1/2N}$. Hence horizon exists for $\Lambda>0$ if $M<1$ and for $\Lambda<0$ if $M>1$. Also the horizon is event if $f'(r_H)>0$ which gives
\begin{align}
\text{Positive branch}&: \text{$k=-1$, $\Lambda>0$ and $M<1$\,,}\nonumber\\
\text{Negative branch}&: \text{$k=1$, $\Lambda<0$ and $M>1$\,,}\nonumber
\end{align}
and we have a cosmological horizon for the opposite $\Lambda$.\\
Notice that the negative branch will always have a branch singularity outside the event horizon. In all other cases we have a smooth spacetime, if no branch singularity. Obviously we assumed $M>0$ in our discussion.

\section{Stability of Pure Lovelock Black Holes}

We turn now to the stability of the black hole derived previously (\ref{eq:BH}). We have

\begin{align}
\label{eq:h}
\frac{h'(r)}{h(r)} =\frac{d-4}{r}&+\frac{M(d-1)}{N(Mr+\Lambda r^d)}-\frac{M(d-1)(d-2N-1)}{M(d-2N-1)r+(d-3)N\Lambda r^d}
\end{align}
while $h'/h=(d-3N-1)/(Nr)$ for $\Lambda=0$ and $d\neq 2N+1$.
Considering the condition $h'/h>0$ we see that we have always instability for the case $\Lambda=0$ in critical even dimensions $d=2N+2$. In the case $d=2N+1$, the last term of (\ref{eq:h}) is zero and we have $h'/h=n/r>0$. Notice that the difference of stability is because the limit when $d\rightarrow 2N+1$ and $\Lambda \rightarrow 0$ do not commute for (\ref{eq:h}).\\
We turn now to the case $\Lambda\neq 0$. 

\subsection{Stability of pure Lovelock in even dimensions}
As we have seen previously, only 3 non trivial solutions are interesting. It is for $\Lambda>0$ and

\begin{align}
\label{eq:1}
f(r)&=1- r^2\Bigl(\frac{M}{r^{2N+1}}+\Lambda\Bigr)^{1/N}\,, ~0<M<M_{\text{ex}}\\
\label{eq:2}
f(r)&=- 1+ r^2\Bigl(\frac{M}{r^{2N+1}}+\Lambda\Bigr)^{1/N}\,, ~0<M<M_{\text{ex}}\\
\label{eq:3}
f(r)&=-1+r^2\Bigl(\frac{M}{r^{2N+1}}+\Lambda\Bigr)^{1/N}\,, ~~M<0
\end{align}
We have stability $(h'/h>0)$ when $r>r_{\text{cr}}$, where $r_{\text{cr}}=c_d (M /\Lambda)^{1/(d-1)}$ is a critical radius and $c_d$ is a coefficient which depends only on the dimension $d$. The spacetime will be stable if $r_e>r_{\text{cr}}$, where $r_e$ is the position of the event horizon.\\
For the first case (\ref{eq:1}) and using $h'/h>0$ along with (\ref{eq:h}) we want $f(r_{\text{cr}})<0$ and $f'(r_{\text{cr}})>0$ which translates to
\begin{align}
\label{stability}
\Lambda M^{d-2}>4^{1-d}\frac{(d-2)^{d-2}}{(d-1)^{d-1}}\frac{\Bigl(6-d+\sqrt{d(d-4)+12}\Bigr)^d}{d^2-2d+6+d\sqrt{d(d-4)+12}}
\end{align}
In the other cases (\ref{eq:2},\ref{eq:3}) we can have $h'/h>0$ but we will always have $h<0$ for the relevant dimensions $d=\{6,10,14,\cdots\}$. Hence these cases are excluded according to the no-ghost condition. Therefore we conclude that the unique stable and ghost free spacetime is the solution (\ref{eq:1}) for which a critical mass (minimum value) can be defined. The stability is guaranteed if $M>M_{\text{cr}}$ where $M_{\text{cr}}$ is easily derived from (\ref{stability}).

\subsection{Stability of pure Lovelock in odd dimensions}

Considering only the case $\Lambda>0$ for which we do not have a branch singularity, we have 
\begin{align}
\frac{h'}{h}=\frac{M(2N-1)+(2N-3)\Lambda r^{2N}}{r(M+\Lambda r^{2N})}
\end{align}
which is positive iff $-\Lambda r^{2N}/M<(2N-1)/(2N-3)$. This is always true ($N>1$). 
Also we have $h\propto \Lambda/\psi>0$ if $\psi>0$ (negative branch). Therefore the only ghost free stable solution under tensor perturbations is 
\begin{align}
f=1-r^2\Bigl(\frac{M}{r^{2N}}+\Lambda\Bigr)^{1/N}
\end{align}
which has a cosmological horizon. This means there exists no stable pure Lovelock black hole in odd dimensions which is however understandable as potential due to mass is constant.

\section{Dimensionally continued BTZ black holes}

In this section, we consider a constraint on the parameters of the model. In this case, we have \cite{Banados:1993ur} $\hat \alpha_m = \hat \alpha_N {N \choose m} ~\Lambda^{2(N-m)}$. Which gives from (\ref{eq:psi})
\begin{align}
\sum_{m=0}^N\hat \alpha_m \psi^m &= \hat \alpha_N \sum_{m=0}^N {N \choose m} ~\Lambda^{2(N-m)} \psi^m= \hat \alpha_N (\Lambda^{2}+\psi)^N=\frac{\mu}{r^{d-1}}
\end{align}
Hence we have
\begin{align}
f=k-\frac{M^{1/N}}{r^{(d-1-2N)/N}}+\Lambda^2 r^2\,,
\end{align}
where $M=\mu/\hat \alpha_N$. Notice that in the special case of odd dimensions $d=2N+1$, we have 
\begin{align}
f=k-M^{2/(d-1)}+\Lambda^2 r^2\,.
\end{align}
The existence of a horizon demands $M^{2/(d-1)}>k$. 

In the general case, we have after some algebra
\begin{align}
h&=\hat \alpha_N\frac{N}{d-2}r^{d-4}\Bigl[(\Lambda^2+\psi)^{N-1}+\frac{r}{n-1}\frac{d}{dr} (\Lambda^2+\psi)^{N-1}\Bigr]\,,\nonumber\\
&=\hat \alpha_N\frac{(d-1-2N)}{(d-2)(d-3)}\frac{M^{(N-1)/N}}{r^{(3N-d+1)/N}}
\end{align}
Hence we have instability for even dimensions $(h'<0)$. But for critical odd dimensions, we have $h=0$. This is simply because potential 
due to mass is constant and that is why there is no central singularity. It is therefore neutral to perturbations. \\
More generically, we will have $h=0$ (from (\ref{eq:A})) iff
\begin{align}
\label{eq:hegal0}
\sum_{m=1}^N m \hat \alpha_m \psi^{m-1}=\frac{\alpha}{r^{d-3}}
\end{align}
where $\alpha$ is a constant, along with the equation for $\psi$

\begin{align}
\sum_{m=0}^N \hat \alpha_m \psi^{m}=\frac{\mu}{r^{d-1}}
\end{align}
Differentiating the last equation wrt $\psi$ and using (\ref{eq:hegal0}), we have
\begin{align}
\frac{\alpha}{r^{d-3}}=-\frac{\mu(d-1)}{\psi' r^d}
\end{align}
which gives $\psi=\beta+\frac{\mu(d-1)}{2\alpha r^2}$, hence the unique black hole for which $h=0$ is of the form
\begin{align}
f=a_1+a_2 r^2
\end{align}
where $(a_1,a_2)$ are 2 constants.

\section{Einstein vs EGB vs pure GB}

Considering the theory in 6D, we have for $L=\alpha_1 R+\alpha_2 R_{GB}-2 \lambda$
\begin{align}
h=r^2\Bigl(\alpha_1+4\alpha_2(3\psi+r\psi')\Bigr)
\end{align}
In the case of Einstein, we have always stability if $\alpha_1>0$, in fact $h=\alpha_1 r^2>0$ and $h'>0$. The case of pure GB has been discussed previously. In fact $\psi=(\frac{M}{r^5}+\Lambda)^{1/2}$, where $\Lambda=\lambda/60\alpha_2$
\begin{align}
h&=\frac{2\alpha_2}{r^3 \psi}\Bigl(M+6 \Lambda r^5\Bigr)\,,\\
h'&=\alpha_2 r \psi \Bigl(24-\frac{25M^2}{(M+\Lambda r^5)^2}\Bigr)
\end{align}
We have stability  if $r>r_{cr}=((5/2\sqrt{6}-1)M/\Lambda)^{1/5}$. In order to have stability of the spacetime we need to impose $r_{cr}<r_e$, which implies  $M^4\Lambda>0.019$.

For the most general case of EGB in 6-d, we have $\psi=-\alpha_1/12\alpha_2\pm (\frac{M}{r^5}+\Lambda+\alpha_1^2/144\alpha_2^2)^{1/2}$, which is already studied in literature (see  \cite{Dotti:2004sh,Dotti:2005sq,Takahashi:2009xh}). It was concluded that there exists critical mass below which the solution becomes unstable. For the particular case of $\Lambda=0$, we have \cite{Takahashi:2010gz}
\begin{align}
M<\frac{2\sqrt{3}(1+\sqrt{10}+\sqrt{15})}{(-1+\sqrt{10}+\sqrt{15})^{3/2}}	 \sqrt{\frac{\alpha_2}{\alpha_1}}\,.
\end{align}
The constraint can be generalized easily, e.g. we have for $\Lambda>0$,
\begin{align}
M<\frac{4\sqrt{3}(12+5\sqrt{6})}{\Bigl(-1+\sqrt{(25+10\sqrt{6})(1+X)}\Bigr)^{5/2}}\Bigl(1+X\Bigr)\sqrt{\frac{\alpha_2}{\alpha_1}}\,,
\end{align}
where $X=144\alpha_2^2\Lambda/\alpha_1^2$.\\
Hence we see that except for the Einstein case, we will always have an instability for small masses. Also the constraint is stronger in Einstein-Gauss-Bonnet compared to pure Gauss-Bonnet if $\Lambda$ is large ($\Lambda>2.6\times 10^{-3} \alpha_1^2/\alpha_2^2$). In the contrary, when $\Lambda$ is small, Einstein-Gauss-Bonnet allows a larger spectrum of stability for the mass.

\section{Conclusion}

In this paper we have derived the master equation for tensor perturbations of black holes in any order Lovelock theory, namely, in any dimensions, by expanding action to second order of perturbations. Hence we have derived no-ghost and tachyonic stability conditions for a spherically symmetric solution in Lovelock gravity. The relevant perturbation defining function $h$ must be positive to avoid a ghost and $h'>0$ to have stability of black hole (tachyonic instability). We apply this stability analysis to pure Lovelock theory in even $d=2N+2$ dimensions. It turns out that pure Lovelock black hole spacetime is always unstable unless $\Lambda$ is brought in. $\Lambda$ therefore plays a stabilizing role for even dimensional pure Lovelock black holes, and stability is achieved by prescribing a lower bound on its 
mass while an upper bound on mass comes from existence of horizon. The latter is derived by using the standard theorems of resultant for 
the polynomial $P=\Lambda r^{d-1}-r+M$. Thus black hole mass is bounded on either ends.  

Intuitively lower and upper mass bounds could perhaps be understood as follows: Without $\Lambda$, pure Lovelock black hole for $N>1$ is always unstable. Note that gravitational potential for pure Lovelock black hole goes as $1/r^{(d-2)/2}$ which is weaker than Einstein potential going as  $1/r^{d-3}$. Now when $\Lambda$ is introduced which implies repulsion going as $r$ and it effectively defines instability threshold radius which could be pushed inside black hole horizon if it is sufficiently massive. This is how the lower mass threshold comes about. The presence of $\Lambda$ would however always define upper threshold for mass in all cases, even for Einstein gravity (see also \cite{Takahashi:2011du}). 

For Einstein gravity, there is always stability and so is also the case for odd dimensional (BTZ-like) pure Lovelock black hole. The latter however has no central singularity, it essentially means that dS-like space is stable under tensor perturbations but it may not be stable for scalar perturbations. In particular, $6$-dimensional pure GB black hole is stable for  $M^4\Lambda>0.019$. There also exists a lower mass bound for stability of E-GB black hole. In the case of dimensionally continued black holes, even dimensional ones are unstable while for odd dimensional ones, $h=0$, and hence they are neutral for tensor perturbations. 

In Einstein-Lovelock theory, there are as many coupling constants as the degree of polynomial $N$ which cannot be determined. Since there is only one force which can determine only one coupling constant. Hence besides $\Lambda$, which is a constant of spacetime structure \cite{d2}, there should only be one free coupling. This is what is done in pure Lovelock gravity while for dimensionally continued case all couplings are related to the unique vacuum defined by $\Lambda$ \cite{Banados:1993ur}. Then dimensionally continued black holes are unstable while pure Lovelock ones could be made stable by $\Lambda$. If stability is the determining criterion, pure Lovelock black holes score over dimensionally continued ones. We thus have a remarkably interesting result for pure Lovelock black holes that existence of horizon and stability bounds its mass between two upper and lower thresholds; i.e. $M_{\text{cr}} < M < M_{\text{ex}}$.

\ack{
RG wants to thank the National Research Foundation and the University of KwaZulu-Natal for financial support. We thank the anonymous referee for constructive criticism.
}

\appendix
\section{First order}

The first order variation of the Riemann tensor is given by

\begin{align}
\delta R_{0i}^{~~0j} &= \Bigl(\frac{\ddot \phi}{2f}-\frac{f'}{4}\phi'\Bigr)\bar h_i^j\,,\\
\delta R_{1i}^{~~1j} &= \Bigl[-f\frac{\phi''}{2}-(\frac{f'}{4}+\frac{f}{r})\phi'\Bigr]\bar h_i^j\,,\\
\delta R_{ij}^{~~kl} &= -\frac{1}{2}\Bigl[\frac{\phi}{r^2}\bar\nabla_{[i}^{~[k}\bar h_{j]}^{~l]}+(\frac{\kappa\phi}{r^2}+\frac{f\phi'}{r})\delta_{[i}^{[k}\bar h_{j]}^{~l]}\Bigr]\,,\\
\delta R_{0i}^{~~1j} &=\Bigl[-\frac{f}{2}\dot\phi'+\Bigl(\frac{f'}{4}-\frac{f}{2r}\Bigr)\dot\phi\Bigr]\bar h_i^j\,,\\
\delta R_{1i}^{~~0j} &=\Bigl[\frac{\dot\phi'}{2f}+\Bigl(\frac{1}{2rf}-\frac{f'}{4f^2}\Bigr)\dot\phi\Bigr]\bar h_i^j\,,\\
\delta R_{0i}^{~~jk} &=\frac{\dot\phi}{2r^2}\Bigl(\bar\nabla^k\bar h_i^j-\bar\nabla^j\bar h_i^k\Bigr)\,,\\
\delta R_{ij}^{~~0k} &=\frac{\dot\phi}{2f}\Bigl(\bar\nabla_i\bar h_j^k-\bar\nabla_j\bar h_i^k\Bigr)\,,\\
\delta R_{1i}^{~~jk} &=\frac{\phi'}{2r^2}\Bigl(\bar\nabla^k\bar h_i^j-\bar\nabla^j\bar h_i^k\Bigr)\,,\\
\delta R_{ij}^{~~1k} &=-\frac{f \phi'}{2}\Bigl(\bar\nabla_i\bar h_j^k-\bar\nabla_j\bar h_i^k\Bigr)\,.
\end{align}

\section{Second order}

The second order variation of the Riemann tensor is given by

\begin{align}
\delta^2 R_{ab}^{~~cd} &=-\frac{1}{2}R_{ab}^{~~e[c}f_g^{d]} f_e^g+\frac{1}{2}f_e^{[d}\nabla_{[a}^{~c]}f_{b]}^e+\frac{1}{2}\nabla_{[a}^{~e}f_{b]}^{[d} f_e^{c]}
+\frac{1}{4}\nabla_e f_{[a}^d \nabla^e f_{b]}^c-\frac{1}{4}\nabla_{[a} f^{e[c} \nabla^{d]} f_{b]e}
\nonumber\\
&+\frac{1}{4}\nabla^c f_{[b}^e \nabla^d f_{a]e}
+\frac{1}{4}\nabla_{[a} f^{ed} \nabla_{b]} f_e^{c}-\frac{1}{4}\nabla^e f_{[b}^{[c} \nabla_{a]} f_e^{d]} +\frac{1}{4}\nabla_e f_{[b}^{[d} \nabla^{c]} f_{a]}^e\,.
\end{align}

which gives for the relevant formulas

\begin{align}
\delta^2 R_{01}^{~~01} &=0\,,\nonumber\\
\delta^2 R_{0i}^{~~0j} &=-\frac{1}{2f}\Bigl[\phi(\ddot \phi-\frac{1}{2}f f' \phi')+\frac{\dot\phi^2}{2}\Bigr]\bar h_i^k \bar h_k^j\,,\\
\delta^2 R_{1i}^{~~1j} &=\Bigl[\frac{f}{2}\phi \phi''+\frac{f}{4}\phi'^2+\frac{f'}{4}\phi \phi'+\frac{f}{r}\phi \phi'\Bigr]\bar h_i^k \bar h_k^j\,,\\
\delta^2 R_{ij}^{~~kl} &=\Bigl(\frac{\dot \phi^2}{4f}-\frac{f}{4}\phi'^2\Bigr)\bar h_{[i}^k \bar h_{j]}^l-\Bigl(\frac{k}{2r^2}\phi^2+\frac{f}{2r}\phi\phi'\Bigr)\bar h_e^{[k} \delta_{[i}^{l]}\bar h_{j]}^e+\frac{\phi^2}{4r^2}\Bigl(\bar \nabla_e \bar h_{[i}^l\bar\nabla^{e}\bar h_{j]}^k
\nonumber\\
&+
\bar \nabla_{[i} \bar h^{e[l}\bar\nabla^{k]}\bar h_{j]e}+
\bar \nabla^{k} \bar h_{[j}^{e}\bar\nabla^{l}\bar h_{i]e}+
\bar \nabla_{[i} \bar h^{el}\bar\nabla_{j]}\bar h_{e}^k+
\bar \nabla^{e} \bar h_{[j}^{[l}\bar\nabla_{i]}\bar h_{e}^{k]}+
\bar \nabla_{e} \bar h_{[j}^{[l}\bar\nabla^{k]}\bar h_{i]}^e
\Bigr)\nonumber\\
&+\frac{\phi^2}{2r^2}\Bigl(\bar h_e^{[l}\bar \nabla_{[i}^{~k]}\bar h_{j]}^e+\bar\nabla_{[i}^{~e}\bar h_{j]}^{[l} \bar h_e^{k]}\Bigr)\,.
\end{align}
\begin{align}
\nonumber
\end{align}
From the last equation, we can derive a useful formula
\begin{align}
\delta^2 R_{ij}^{~~kl}\delta_{kl}^{ij}&= \frac{\phi^2}{2r^2}\Bigl(3\bar \nabla_e \bar h_i^j\bar\nabla^e \bar h_j^i-2\bar \nabla_e \bar h_i^j\bar \nabla^i \bar h_j^e\Bigr)+\Bigl[\frac{f}{2}\phi'^2-\frac{\dot\phi^2}{2f}\nonumber\\
&+2\frac{n-1}{r^2}\Bigl(k\phi^2+rf\phi\phi'\Bigr)+2\frac{\phi^2}{r^2}(\gamma-kn)\Bigr]\bar h_i^j \bar h_j^i\,.
\end{align}


\begin{thebibliography}{99}
\bibitem{lov}
D. ~Lovelock, J. Math. Phys. {\bf 12}, 498 (1971); {\bf 13}, 874 (1972).
\bibitem{Gross:1986iv}
D.~J.~Gross and E.~Witten,
  %``Superstring Modifications of Einstein's Equations,''
  Nucl.\ Phys.\ B {\bf 277} (1986) 1.
\bibitem{Zumino:1985dp}
  B.~Zumino,
  %``Gravity Theories in More Than Four-Dimensions,''
  Phys.\ Rept.\  {\bf 137} (1986) 109.
  %%CITATION = PRPLC,137,109;%%
  %292 citations counted in INSPIRE as of 23 Oct 2013

%\cite{Zwiebach:1985uq}
\bibitem{Zwiebach:1985uq}
  B.~Zwiebach,
  %``Curvature Squared Terms and String Theories,''
  Phys.\ Lett.\ B {\bf 156} (1985) 315.

\bibitem{d1}
N. ~Dadhich, Universalization as a physical guiding principle, arxiv:gr-qc/0311028; Probing universality of gravity, arxiv:gr-qc/0407003. 
\bibitem{d2}
N. ~Dadhich, On the Gauss-Bonnet gravity, arxiv:hep-th/0509126; Pramana {\bf 77}, 443 (2011), [arxiv:1006.1552]; Int. J. Mod. Phys. {\bf D20}, 1739 (2011), [arxiv: 1105.3396]. 
\bibitem{d3}
N. ~Dadhich, On the Gauss-Bonnet gravity, arxiv:hep-th/0509126; Universality, Gravity, enigmatic $\Lambda$ and beyond, arxiv:gr-qc/0405115; Int. J. Mod. Phys. {\bf D20}, 1739 (2011), [arxiv: 1105.3396].

\bibitem{dgj1}
N.~Dadhich, S.~G.~Ghosh and S.~Jhingan,
  %``The Lovelock gravity in the critical spacetime dimension,''
  Phys.\ Lett.\ B {\bf 711}, 196 (2012), [arxiv:1202.4575].
\bibitem{bianchi}
 N. Dadhich, Pramana {\bf74}, 875 (2010) (arXiv:0802.3034)
\bibitem{dpp1}
  N.~Dadhich, J.~M.~Pons and K.~Prabhu,
  %``Thermodynamical universality of the Lovelock black holes,''
  Gen.\ Rel.\ Grav.\  {\bf 44}, 2595 (2012) [arxiv:1110.0673]
\bibitem{dpp2}
  N.~Dadhich, J.~M.~Pons and K.~Prabhu,
  %``Thermodynamical universality of the Lovelock black holes,''
  Gen.\ Rel.\ Grav.\  {\bf 45}, 1131 (2013)
  [arXiv:1201.4994 [gr-qc]].
 \bibitem{prague}
 N.~Dadhich, The gravitational equation in higher dimensions, Relativity and Gravitation: 100 years after Einstein in Prague, June 25-28, (2012) [arXiv:1210.3022].
\bibitem{dgj2}
 N.~Dadhich, S.~G.~Ghosh and S.~Jhingan, Bound orbits and gravitational theory, arxiv:1308.4770v1.
%\cite{Zumino:1985dp}
  %%CITATION = PHLTA,B156,315;%%
  %693 citations counted in INSPIRE as of 23 Oct 2013

%\cite{Hollands:2012xy}
\bibitem{hollands}
  S.~Hollands and A.~Ishibashi,
  %``Black hole uniqueness theorems in higher dimensional spacetimes,''
  Class.\ Quant.\ Grav.\  {\bf 29} (2012) 163001
  [arXiv:1206.1164 [gr-qc]].
  %%CITATION = ARXIV:1206.1164;%%
  %10 citations counted in INSPIRE as of 14 Nov 2013

%\cite{Regge:1957td}
\bibitem{Regge:1957td}
  T.~Regge and J.~A.~Wheeler,
  %``Stability of a Schwarzschild singularity,''
  Phys.\ Rev.\  {\bf 108} (1957) 1063.
  %%CITATION = PHRVA,108,1063;%%
  %855 citations counted in INSPIRE as of 23 Oct 2013

%\cite{Zerilli:1970se}
\bibitem{Zerilli:1970se}
  F.~J.~Zerilli,
  %``Effective potential for even parity Regge-Wheeler gravitational perturbation equations,''
  Phys.\ Rev.\ Lett.\  {\bf 24} (1970) 737.
  %%CITATION = PRLTA,24,737;%%
  %309 citations counted in INSPIRE as of 23 Oct 2013

%\cite{Vishveshwara:1970cc}
\bibitem{Vishveshwara:1970cc}
  C.~V.~Vishveshwara,
  %``Stability of the schwarzschild metric,''
  Phys.\ Rev.\ D {\bf 1} (1970) 2870.
  %%CITATION = PHRVA,D1,2870;%%
  %201 citations counted in INSPIRE as of 23 Oct 2013

%\cite{Price:1971fb}
\bibitem{Price:1971fb}
  R.~H.~Price,
  %``Nonspherical perturbations of relativistic gravitational collapse. 1. Scalar and gravitational perturbations,''
  Phys.\ Rev.\ D {\bf 5} (1972) 2419.
  %%CITATION = PHRVA,D5,2419;%%
  %502 citations counted in INSPIRE as of 25 Oct 2013

\bibitem{Wald:1979}
R.~M.~Wald,
J.\ Math.\ Phys.\ {\bf 20} (1979), 1056;
J.\ Math.\ Phys.\ {\bf 21} (1980), 218.

%\cite{Gibbons:2002pq}
\bibitem{Gibbons:2002pq}
  G.~Gibbons and S.~A.~Hartnoll,
  %``A Gravitational instability in higher dimensions,''
  Phys.\ Rev.\ D {\bf 66} (2002) 064024
  [hep-th/0206202].
  %%CITATION = HEP-TH/0206202;%%
  %122 citations counted in INSPIRE as of 25 Oct 2013

%\cite{Ishibashi:2003ap}
\bibitem{Ishibashi:2003ap}
  A.~Ishibashi and H.~Kodama,
  %``Stability of higher dimensional Schwarzschild black holes,''
  Prog.\ Theor.\ Phys.\  {\bf 110} (2003) 901
  [hep-th/0305185].
  %%CITATION = HEP-TH/0305185;%%
  %165 citations counted in INSPIRE as of 18 Aug 2013


%\cite{Kodama:2003jz}
\bibitem{Kodama:2003jz}
  H.~Kodama and A.~Ishibashi,
  %``A Master equation for gravitational perturbations of maximally symmetric black holes in higher dimensions,''
  Prog.\ Theor.\ Phys.\  {\bf 110} (2003) 701
  [hep-th/0305147].
  %%CITATION = HEP-TH/0305147;%%
  %202 citations counted in INSPIRE as of 25 Oct 2013


%\cite{Gleiser:2005ra}
\bibitem{Gleiser:2005ra}
  R.~J.~Gleiser and G.~Dotti,
  %``Linear stability of Einstein-Gauss-Bonnet static spacetimes. Part II: Vector and scalar perturbations,''
  Phys.\ Rev.\ D {\bf 72} (2005) 124002
  [gr-qc/0510069].
  %%CITATION = GR-QC/0510069;%%
  %65 citations counted in INSPIRE as of 25 Oct 2013

%\cite{Takahashi:2010ye}
\bibitem{Takahashi:2010ye}
  T.~Takahashi and J.~Soda,
  %``Master Equations for Gravitational Perturbations of Static Lovelock Black Holes in Higher Dimensions,''
  Prog.\ Theor.\ Phys.\  {\bf 124} (2010) 911
  [arXiv:1008.1385 [gr-qc]].
  %%CITATION = ARXIV:1008.1385;%%
  %15 citations counted in INSPIRE as of 25 Oct 2013

\bibitem{Dotti:2004sh}
  G.~Dotti and R.~J.~Gleiser,
  %``Gravitational instability of Einstein-Gauss-Bonnet black holes under tensor mode perturbations,''
  Class.\ Quant.\ Grav.\  {\bf 22} (2005) L1
  [gr-qc/0409005].
  %%CITATION = GR-QC/0409005;%%
  %56 citations counted in INSPIRE as of 25 Oct 2013

%\cite{Dotti:2005sq}
\bibitem{Dotti:2005sq}
  G.~Dotti and R.~J.~Gleiser,
  %``Linear stability of Einstein-Gauss-Bonnet static spacetimes. Part I. Tensor perturbations,''
  Phys.\ Rev.\ D {\bf 72} (2005) 044018
  [gr-qc/0503117].
  %%CITATION = GR-QC/0503117;%%
  %59 citations counted in INSPIRE as of 25 Oct 2013

%\cite{Takahashi:2009dz}
\bibitem{Takahashi:2009dz}
  T.~Takahashi and J.~Soda,
  %``Stability of Lovelock Black Holes under Tensor Perturbations,''
  Phys.\ Rev.\ D {\bf 79} (2009) 104025
  [arXiv:0902.2921 [gr-qc]].
  %%CITATION = ARXIV:0902.2921;%%
  %33 citations counted in INSPIRE as of 25 Oct 2013

%\cite{Takahashi:2009xh}
\bibitem{Takahashi:2009xh}
  T.~Takahashi and J.~Soda,
  %``Instability of Small Lovelock Black Holes in Even-dimensions,''
  Phys.\ Rev.\ D {\bf 80} (2009) 104021
  [arXiv:0907.0556 [gr-qc]].
  %%CITATION = ARXIV:0907.0556;%%
  %22 citations counted in INSPIRE as of 25 Oct 2013

%\cite{Takahashi:2010gz}
\bibitem{Takahashi:2010gz}
  T.~Takahashi and J.~Soda,
  %``Catastrophic Instability of Small Lovelock Black Holes,''
  Prog.\ Theor.\ Phys.\  {\bf 124} (2010) 711
  [arXiv:1008.1618 [gr-qc]].
  %%CITATION = ARXIV:1008.1618;%%
  %19 citations counted in INSPIRE as of 25 Oct 2013

\bibitem{btz}
M.~Banados, C.~Teitelboim and J.~Zanelli, Phys. Rev. Lett. {\bf 69}, 1849 (1992)
%\cite{Dotti:2004sh}


%\cite{Canfora:2010rh}
\bibitem{Canfora:2010rh}
  F.~Canfora and A.~Giacomini,
  %``BTZ-like black holes in even dimensional Lovelock theories,''
  Phys.\ Rev.\ D {\bf 82} (2010) 024022
  [arXiv:1005.0091 [gr-qc]].
  %%CITATION = ARXIV:1005.0091;%%
  %6 citations counted in INSPIRE as of 26 May 2014

%\cite{Anabalon:2011bw}
\bibitem{Anabalon:2011bw}
  A.~Anabalon, F.~Canfora, A.~Giacomini and J.~Oliva,
  %``Black holes with gravitational hair in higher dimensions,''
  Phys.\ Rev.\ D {\bf 84} (2011) 084015
  [arXiv:1108.1476 [hep-th]].
  %%CITATION = ARXIV:1108.1476;%%
  %5 citations counted in INSPIRE as of 26 May 2014

%\cite{Boulware:1985wk}
\bibitem{Boulware:1985wk}
  D.~G.~Boulware and S.~Deser,
  %``String Generated Gravity Models,''
  Phys.\ Rev.\ Lett.\  {\bf 55} (1985) 2656.
  %%CITATION = PRLTA,55,2656;%%
  %629 citations counted in INSPIRE as of 14 Nov 2013

%\cite{Wheeler:1985qd}
\bibitem{Wheeler:1985qd}
  J.~T.~Wheeler,
  %``Symmetric Solutions to the Maximally {Gauss-Bonnet} Extended Einstein Equations,''
  Nucl.\ Phys.\ B {\bf 273} (1986) 732.
  %%CITATION = NUPHA,B273,732;%%
  %177 citations counted in INSPIRE as of 14 Nov 2013

%\cite{Deser:2005pc}
\bibitem{Deser:2005pc}
  S.~Deser and A.~V.~Ryzhov,
  %``Curvature invariants of static spherically symmetric geometries,''
  Class.\ Quant.\ Grav.\  {\bf 22} (2005) 3315
  [gr-qc/0505039].
  %%CITATION = GR-QC/0505039;%%
  %10 citations counted in INSPIRE as of 14 Nov 2013

%\cite{Higuchi:1986wu}
\bibitem{Higuchi:1986wu}
  A.~Higuchi,
  %``Symmetric Tensor Spherical Harmonics On The N Sphere And Their Application To The De Sitter Group So(n,1),''
  J.\ Math.\ Phys.\  {\bf 28} (1987) 1553
   [Erratum-ibid.\  {\bf 43} (2002) 6385].
  %%CITATION = JMAPA,28,1553;%%
  %91 citations counted in INSPIRE as of 18 Aug 2013

\bibitem{Buell}
  W.~F.~Buell and B.~A.~Shadwick,
  %``Potentials and bound states,''
Am. J. Phys. 63, 256. (1995)

%\cite{Moncrief:1974am}
\bibitem{Moncrief:1974am}
  V.~Moncrief,
  %``Gravitational perturbations of spherically symmetric systems. I. The exterior problem.,''
  Annals Phys.\  {\bf 88} (1974) 323.
  %%CITATION = APNYA,88,323;%%
  %216 citations counted in INSPIRE as of 13 Jun 2014

%\cite{DeFelice:2011ka}
\bibitem{DeFelice:2011ka}
  A.~De Felice, T.~Suyama and T.~Tanaka,
  %``Stability of Schwarzschild-like solutions in f(R,G) gravity models,''
  Phys.\ Rev.\ D {\bf 83} (2011) 104035
  [arXiv:1102.1521 [gr-qc]].
  %%CITATION = ARXIV:1102.1521;%%
  %13 citations counted in INSPIRE as of 13 Jun 2014

%\cite{Maldacena:2002vr}
\bibitem{Maldacena:2002vr}
  J.~M.~Maldacena,
  %``Non-Gaussian features of primordial fluctuations in single field inflationary models,''
  JHEP {\bf 0305} (2003) 013
  [astro-ph/0210603].
  %%CITATION = ASTRO-PH/0210603;%%
  %1165 citations counted in INSPIRE as of 13 Jun 2014

%\cite{Cai:2006pq}
\bibitem{Cai:2006pq}
  R.~-G.~Cai and N.~Ohta,
  %``Black Holes in Pure Lovelock Gravities,''
  Phys.\ Rev.\ D {\bf 74} (2006) 064001
  [hep-th/0604088].
  %%CITATION = HEP-TH/0604088;%%
  %36 citations counted in INSPIRE as of 14 Nov 2013

%\cite{Torii:2005xu}
\bibitem{Torii:2005xu}
  T.~Torii and H.~Maeda,
  %``Spacetime structure of static solutions in Gauss-Bonnet gravity: Neutral case,''
  Phys.\ Rev.\ D {\bf 71} (2005) 124002
  [hep-th/0504127].
  %%CITATION = HEP-TH/0504127;%%
  %61 citations counted in INSPIRE as of 26 May 2014

%\cite{Banados:1993ur}
\bibitem{Banados:1993ur}
  M.~Banados, C.~Teitelboim and J.~Zanelli,
  %``Dimensionally continued black holes,''
  Phys.\ Rev.\ D {\bf 49} (1994) 975
  [gr-qc/9307033].
  %%CITATION = GR-QC/9307033;%%
  %153 citations counted in INSPIRE as of 08 Nov 2013


%\cite{Takahashi:2011du}
\bibitem{Takahashi:2011du}
  T.~Takahashi and J.~Soda,
  %``Pathologies in Lovelock AdS Black Branes and AdS/CFT,''
  Class.\ Quant.\ Grav.\  {\bf 29} (2012) 035008
  [arXiv:1108.5041 [hep-th]].
  %%CITATION = ARXIV:1108.5041;%%
  %6 citations counted in INSPIRE as of 04 Mar 2014
\end{thebibliography}
\end{document}